\documentclass[a4paper,11pt]{article}
\usepackage{pos}
\usepackage{listings}
\usepackage{enumitem}
\usepackage{setspace}
\usepackage{multicol}

\definecolor{codebg}{RGB}{240,240,240}

\lstdefinestyle{custom}{
    backgroundcolor=\color{codebg},
    basicstyle=\ttfamily\small,
    numbers=left,
    numberstyle=\tiny,
    breaklines=true,
    frame=single
}

\title{Light and strange quark masses with $N_{f}=2+1$ Wilson fermions}

\author[a,b]{Gregorio Herdoíza}
\author*[a]{Fernando P. Panadero}
\author[a,b]{Carlos Pena}
\author[c]{Alejandro Sáez}

\affiliation[a]{Instituto de Física Teórica UAM-CSIC, c/ Nicolás Cabrera 13-15}
\affiliation[b]{Department of Theoretical Physics, Universidad Autónoma de Madrid, 28049 Madrid, Spain}
\affiliation[c]{Instituto de Física Corpuscular (IFIC), CSIC-Universitat de València, 46071, Valencia, Spain}

\emailAdd{fernando.p@csic.es}

\abstract{We report on the status of an update of our collaboration's previous computation \cite{Bruno:2019vup} of light and strange quark masses in QCD with $N_{f}=2+1$ dynamical flavours. Bare quark masses are extracted from CLS ensembles, using $O(a)$-improved Wilson fermions, and the mass renormalization is performed non-perturbatively in the Schrödinger functional scheme over a wide range of scales to make safe contact with perturbation theory. Results for five lattice spacings, down to $a\sim 0.038 \textrm{ fm}$, and pion masses reaching the physical value are included in the analysis. This allows for the exploration of different models for cutoff and chiral effects, and a controlled extrapolation to the physical point.

\vspace*{0.5cm}
\begin{flushright}
    IFT-UAM/CSIC-26-41 \\
\end{flushright}
}

\FullConference{The 42nd International Symposium on Lattice Field Theory (LATTICE2025)\\
2-8 November 2025\\
Tata Institute of Fundamental Research, Mumbai, India\\}

\begin{document}
\maketitle

\section{Introduction}

Quark masses are fundamental parameters of the Standard Model. %
A first-principles determination of these quantities requires a consistent non-perturbative formulation of QCD. %
We present a preliminary update on \cite{Bruno:2019vup} for the determination of the light and strange quark masses, in a framework where the up and down quarks are mass-degenerate and QED effects are ignored, known as $N_{f} = 2+1$ QCD. %
The status for the determination of quark masses in Lattice QCD with this setup is summarized in the Flavour Lattice Averaging Group (FLAG) Review \cite{FLAG}. %
The current values for the average are
\begin{equation}
  \overline{m}_{ud}^{\rm \scriptscriptstyle \overline{MS}}(2\textrm{GeV}) = 3.387(39) \textrm{MeV}, \qquad \overline{m}_{s}^{\overline{\rm \scriptscriptstyle MS}}(2\textrm{GeV}) = 92.4(1.0) \textrm{MeV}.
\end{equation}

To determine the physical values of quark masses, we employ large-scale simulations with $N_{f} = 2+1$ dynamical flavours produced by the Coordinated Lattice Simulations (CLS) effort \cite{Bruno:2014jqa}. %
These ensembles employ a tree-level Symanzik-improved gauge action and non-perturbatively improved Wilson fermions in the sea sector. %
We consider a set of ensembles with five lattice spacings in the range $0.085\text{-}0.038 \textrm{ fm}$ and pion masses from $420 \textrm{ MeV}$ down to the physical pion mass.

The quark masses extracted from these simulations are bare quantities and have to be renormalized. %
For this, we use the non-perturbative renormalization and Renormalization Group (RG) running results from the ALPHA Collaboration \cite{Campos:2018ahf}, based on a Schrödinger functional (SF) finite volume renormalization scheme. %
This allows to run from a hadronic scale, where the quark masses obtained in the large volume simulations are extracted and renormalized, to a very high energy perturbative scale, where one can make safe contact with perturbation theory.

The main update with respect to \cite{Bruno:2019vup} is the extension of the dataset with six new ensembles that include one additional lattice spacing and physical kinematics ensembles, allowing for a better controlled chiral-continuum extrapolation to the physical point, as well as a new scale setting determination \cite{Bussone:2025wlf}.

\section{Lattice setup and ensembles}

In this section, we summarize the theoretical framework used to define renormalized quark masses and give some details of the ensembles entering the analysis.

\subsection{Theoretical framework}

The main observables for the computation of quark masses are the pseudoscalar-pseudoscalar and axial-pseudoscalar bare correlation functions

\begin{equation}
  f_{P}^{ij}(x_{0},y_{0}) = -\frac{a^{6}}{L^{3}}\sum_{\vec{x},\vec{y}} \langle P^{ij}(x_{0},\vec{x})P^{ji}(y_{0},\vec{y})\rangle, \qquad
  f_{A}^{ij}(x_{0},y_{0}) = -\frac{a^{6}}{L^{3}}\sum_{\vec{x},\vec{y}} \langle A_{0}^{ij}(x_{0},\vec{x})P^{ji}(y_{0},\vec{y})\rangle,
\end{equation}
where $A_{\mu}^{ij}(x) ={\overline\psi}^{i}(x)\gamma_{\mu}\gamma_{5}\psi^{j}(x)$, $P^{ij} (x)={\overline\psi}^{i}(x)\gamma_{5}\psi^{j}(x)$ and $i, j = 1,2,3 $ label quark flavour. By imposing the PCAC Ward identity

\begin{equation}
  \partial_{\mu} A_{\mu}^{ij}(x) = 2m_{ij}P^{ij}(x) = (m_{i} + m_{j})P^{ij}(x), \qquad \text{with}\qquad i\neq j,
\end{equation}
we can define the bare PCAC quark mass by using the pseudoscalar density to probe this relation. %
\begin{equation}
  m_{ij}(x_{0}) = \frac{\frac{1}{2}(\partial_{0} + \partial^{*}_{0}) f_{A}^{ij}(x_{0},y_{0}) + c_{A} a \partial_{0}\partial^{*}_{0}f^{ij}_{P}(x_{0},y_{0})}{2f_{P}^{ij}(x_{0},y_{0})},
  \label{eq:plat_mq}
\end{equation}
where the mass independent improvement coefficient $c_{A}$ is non-perturbatively determined in \cite{Bulava:2015bxa}. The renormalized quark masses are then defined as

\begin{equation}
  a\overline{m}_{ij}(g_{0},a\mu) = \frac{Z_{\rm \scriptscriptstyle A}(g_{0}^{2})}{Z_{\rm \scriptscriptstyle P}(g_{0}^{2},a\mu)} am_{ij}\left[1 + (\tilde{b}_{A} - \tilde{b}_{P}) am_{ij}\right],
\end{equation}
where the renormalization factors $Z_{\rm \scriptscriptstyle P}(g_{0},a\mu)$ \cite{Campos:2018ahf}, $Z_{\rm \scriptscriptstyle A}(g_{0}^{2})$ \cite{DallaBrida:2018tpn} and the improvement coefficient $(\tilde{b}_{\rm \scriptscriptstyle A} - \tilde{b}_{\rm \scriptscriptstyle P})$ \cite{deDivitiis:2019xla} have been computed in previous works. %
In this expression, we are neglecting improvement terms proportional to $\textrm{Tr}(aM_{q})$. %
These terms are expected to have a small effect with our current precision as they are $\textrm{O}(g_{0}^{4})$ in perturbation theory, see \cite{Bussone:2025wlf} for more details.

After the chiral-continuum extrapolation, this expresion yields the renormalized value of the quark masses in the Schrödinger Functional scheme at $\mu_{\rm \scriptscriptstyle had}=233(8) \rm{ MeV}$. %
We can then convert this to the RGI masses with the RG factor \cite{Campos:2018ahf}

\begin{equation}
  \frac{M_{i}^{\rm \scriptscriptstyle RGI}}{\overline{m}_{i}(\mu_{\rm \scriptscriptstyle had})} = 0.9148(88).
  \label{eq:RGI_mass}
\end{equation}

As this is an RGI and scheme-independent quantity, it can later be used to extract the quark mass in any other renormalization scheme via

\begin{equation}
  M^{\rm \scriptscriptstyle RGI}_{i} = \overline{m}_{i}(\mu) \left( 2b_{0}\overline{g}^{2}(\mu) \right)^{-\frac{d_{0}}{2b_{0}}} \exp\left\{ -\int_{0}^{\overline{g}(\mu)}dg \left( \frac{\tau(g)}{\beta(g)} - \frac{d_{0}}{b_{0}g}\right)\right\}.
\end{equation}

Where $\beta(g),\tau(g)$ are respectively the beta function and the quark mass anomalous dimension.
In practice, we use the flow scale $t_{0}$ \cite{Luscher:2010iy} to scale both hadron and quark masses:

\begin{equation}
  \phi_{2} = 8t_{0}m_{\pi}^{2}, \qquad \phi_{4} = 8t_{0}(m_{K}^{2} + \frac{1}{2}m_{\pi}^{2}), \qquad \phi_{K} = 8t_{0}m_{K}^{2}
\end{equation}
\begin{equation}
  \phi_{12} = \sqrt{8t_{0}}\text{  } \overline{m}_{12}, \qquad \phi_{13} = \sqrt{8t_{0}}\text{  }  \overline{m}_{13}.
\end{equation}

\subsection{Simulation Setup}

We employ a subset of the $N_{f} = 2+1$ CLS ensembles \cite{Bruno:2014jqa}, generated with $O(a)$-improved Wilson fermion and Lüscher-Weisz action. %
We use the $\textrm{Tr}(M_{q}) = \textrm{const}$ trajectory of CLS, ensuring the lattice spacing is kept fixed, up to $O(a^{2})$, for a fixed value of the bare coupling. A list of the ensembles is given in Table \ref{table:ens_list}.

\begin{table}
  \centering
  \begin{tabular}{ccccccccc}
    Name & $\beta$ & $a$[fm] & $T/a$ & $L/a$ & $m_{\pi}[\textrm{MeV}]$ & $m_{K}[\textrm{MeV}]$ & $m_{\pi}L$ & $N_{cnfg}$ \\  \hline
    H101 & 3.4  & 0.085  & 96 &  32 & 426 & 426 & 5.8 & 2010 \\
    H102 &      &        & 96 &  32 & 360 & 446 & 4.9 & 2005 \\
    H105 &      &        & 96 &  32 & 286 & 470 & 3.9 & 2826 \\  \hline
    H400 & 3.46 & 0.075  & 96 &  32 & 429 & 429 & 5.2 & 1045 \\
    D450 &      &        & 128 & 64 & 221 & 483 & 5.4 & 250 \\  \hline
    N202 & 3.55 & 0.063  & 128 & 48 & 418 & 418 & 6.5 & 899 \\
    N203 &      &        & 128 & 48 & 350 & 448 & 5.4 & 1543 \\
    N200 &      &        & 128 & 48 & 288 & 470 & 4.4 & 1712 \\
    D200 &      &        & 128 & 64 & 204 & 488 & 4.2 & 2001 \\
    E250 &      &        & 192 & 96 & 131 & 497 & 4.0 & 100 \\  \hline
    N300 & 3.70 & 0.049  & 128 & 48 & 427 & 427 & 5.1 & 1521 \\
    N302 &      &        & 128 & 48 & 350 & 457 & 4.2 & 2201 \\
    J303 &      &        & 192 & 64 & 261 & 481 & 4.1 & 1073\\
    E300 &      &        & 192 & 96 & 177 & 499 & 4.2 & 227 \\  \hline
    J500 & 3.85 & 0.038  & 192 & 64 & 418 & 418 & 5.2 & 422 \\
    J501 &      &        & 192 & 64 & 339 & 453 & 4.3 & 787 \\  \hline
  \end{tabular}
  \caption{Set of $N_{f}=2+1$ ensembles entering this analysis. All ensembles have open boundary conditions in the time direction, with the exception of E250 and D450 that have (anti-)periodic boundary conditions. Pion and kaon masses and lattice spacings in physical units are approximate values, see \cite{Bussone:2025wlf} for details.}
  \label{table:ens_list}
\end{table}

We follow the same strategy as in Refs. \cite{Bussone:2025wlf, Bruno:2016plf} to fix a constant value of $\phi_{4}$, computing the mass derivative of primary observables with respect to quark masses and shifting them to a target value of $\phi_{4}$. %
The physical values of $\phi_{2}$ and $\phi_{4}$ are determined in \cite{Bussone:2025wlf} using the so-called Edinburgh Consensus to define isoQCD \cite{FLAG}. %

\begin{figure}
  \includegraphics[scale=0.38]{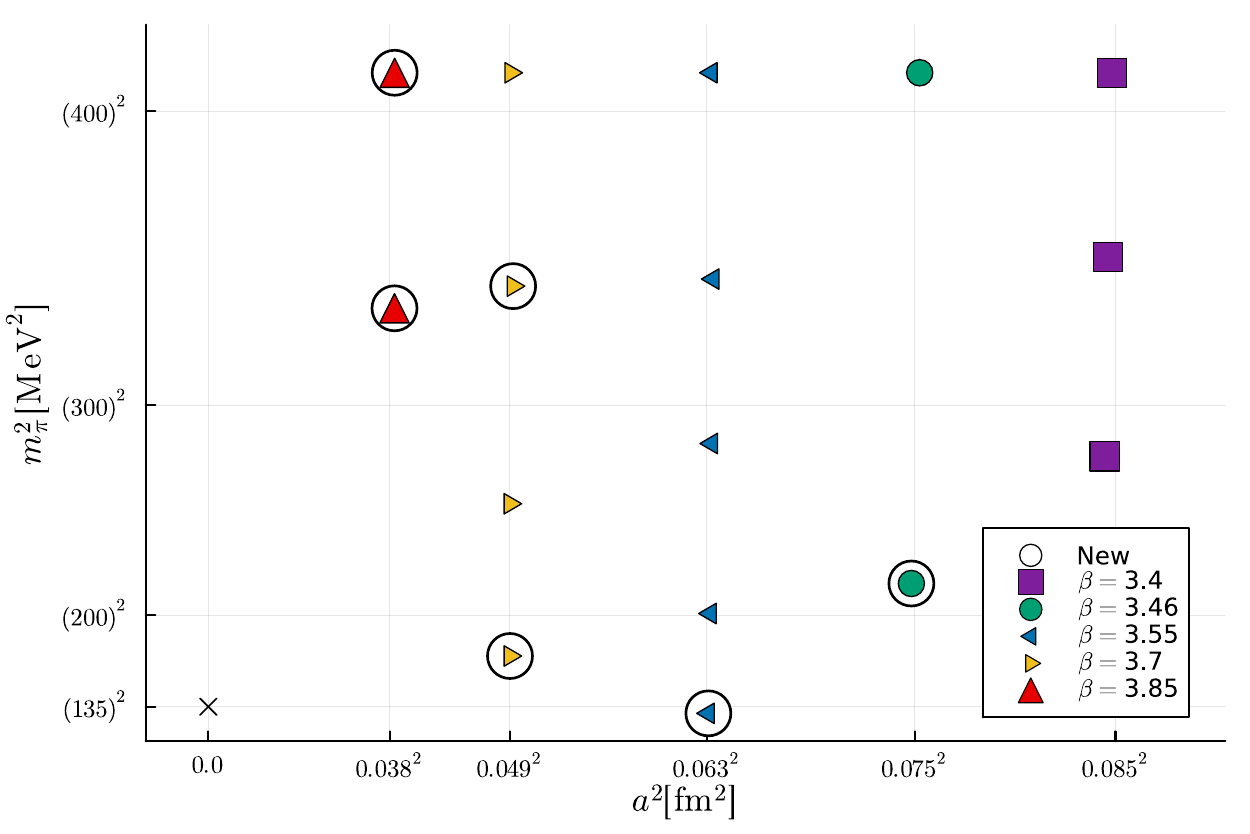}
  \includegraphics[scale=0.38]{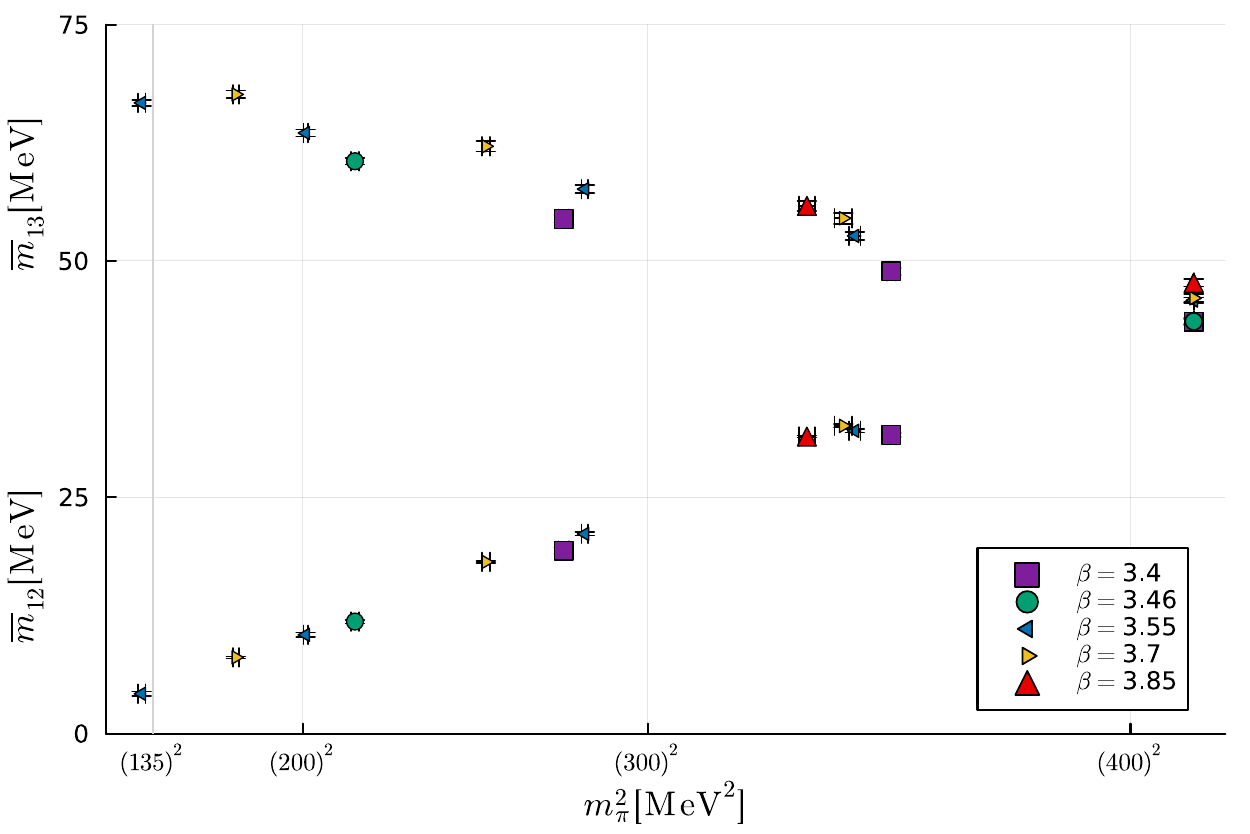}
  \caption{Left: values of the lattice spacing and pion mass for the ensembles used in this work. The new ensembles with respect to \cite{Bruno:2019vup} are highlighted. Right: Values of the light-light ($\overline{m}_{12}$) and light-strange ($\overline{m}_{13}$) renormalized quark masses at $\mu_{had}$ as a function of the pion mass for the same ensembles.}
  \label{fig:cls_data}
\end{figure}

In figure \ref{fig:cls_data} we illustrate the ensembles used in this analysis as a function of the lattice spacing and the pion mass. %
Note that, because of the chosen chiral trajectory of constant $\textrm{Tr}(M_{q})$, the kaon has the same mass as the pion at the $SU(3)-$symmetric point, where the pion is heavier, and becomes heavier at lighter pion masses. %
This feature of the chiral trajectory can also be seen in the right plot of figure \ref{fig:cls_data}, where we show the value for the renormalized average quark masses as a function of the pion mass for various lattice spacings.

\section{Chiral continuum extrapolation}

\subsection{Chiral dependence}

To explore the chiral extrapolation, we perform fits with different functional form dependencies on the pion mass at zero lattice spacing. %

\begin{itemize}
  \item \texttt{[Tay n]} : Taylor expansion of quark masses up to the n-th order in the pion mass around the symmetric point: %
        \begin{align}
          &\phi_{12} = p_{0} + p_{1}(\phi_{2} - \phi_{2}^{sym}) + ... + p_{n}(\phi_{2}-\phi_{2}^{sym})^{n}, \\
          &\phi_{13} = p_{0} + p_{n+1}(\phi_{2} - \phi_{2}^{sym}) + ... + p_{2n}(\phi_{2}-\phi_{2}^{sym})^{n}.
        \end{align}
        Notice that, since both quark masses are exactly the same at the symmetric point for every lattice spacing, the parameter $p_{0}$ is the same in both functional forms.
        While these fits have the advantage of imposing this constraint, there is no extra information about the chiral behaviour for the quark masses.

  \item \texttt{[$\chi$ptm]}: Inspired by next-to-leading order SU(3)-$\chi$PT we can parametrize the chiral dependence as:
        \begin{align}
          &\phi_{12} = \phi_{2} \left(p_{1} + p_{2} \phi_{2} + p_{3}K(\mathcal{L}_{2} - \frac{1}{3}\mathcal{L}_{\eta}) \right), \\
          &\phi_{13} = \phi_{K}\left(p_{1} + p_{2}\phi_{K} + \frac{2}{3}p_{3}K\mathcal{L}_{\eta}\right),
        \end{align}

        where $\mathcal{L}_{X} = \phi_{X}\textrm{log}\phi_{X}$ are the so-called chiral-logs, $K = \left[ 8t_{0} 16\pi^{2} (\frac{1}{3}f_{\pi} + \frac{2}{3}f_{K})^{2} \right]^{-1}$ and $\phi_{\eta} = (4\phi_{4} - 3\phi_{2})/3$. These fits have the advantage of having a reduced number of free parameters, as they are constrained by a physical interpretation, and also provide information about the approach to the chiral point.

  \item \texttt{[$\chi$ptr]}: Instead of fitting the masses directly, we fit the ratios:
        \begin{align}
          &\frac{\phi_{12}}{\phi_{13}} = \frac{2\phi_{2}}{2\phi_{4} - \phi_{2}}\left[ 1 + \frac{p_{2}}{p_{1}} \left( \frac{3}{2}\phi_{2} - \phi_{4}\right) - K(\mathcal{L}_{2} - \mathcal{L}_{\eta})\right]\label{eq:ratio_1},\\
          &\frac{4\phi_{13}}{2\phi_{4}-\phi_{2}} + \frac{\phi_{12}}{\phi_{2}} = 3p_{1} + 2p_{2}\phi_{4} + p_{3}K(\mathcal{L}_{2} + \mathcal{L}_{\eta})\label{eq:ratio2}.
        \end{align}

        One useful feature of this fit ansatz is that chiral continuum effects are somewhat split between the two ratios. The first ratio is constructed to have cutoff effects highly suppressed and will be very sensitive to chiral effects. Meanwhile, the second has chiral effects in $\phi_{2}$ largely suppressed (only the chiral logs remain) and is more sensitive to cutoff effects.

\end{itemize}

\begin{figure}[htbp]
  \centering
  \includegraphics[scale=0.35]{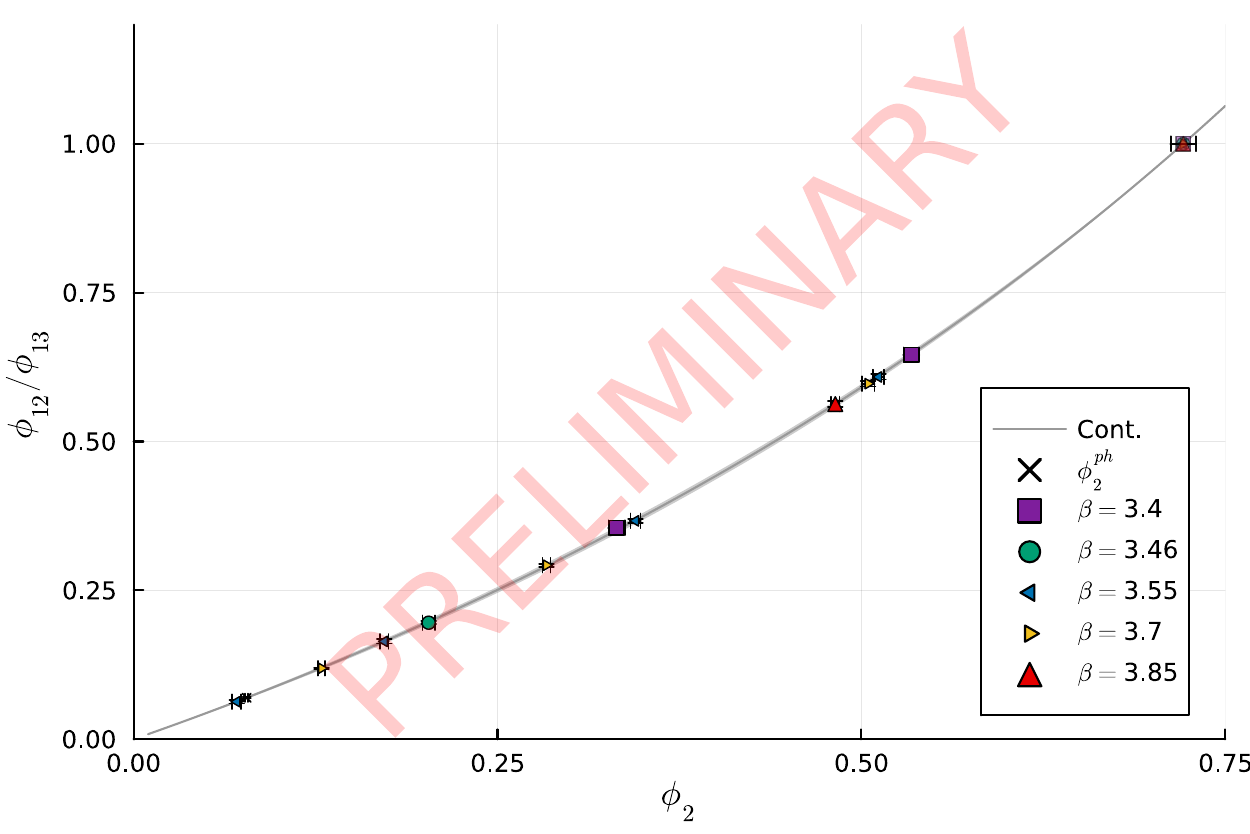}
  \includegraphics[scale=0.35]{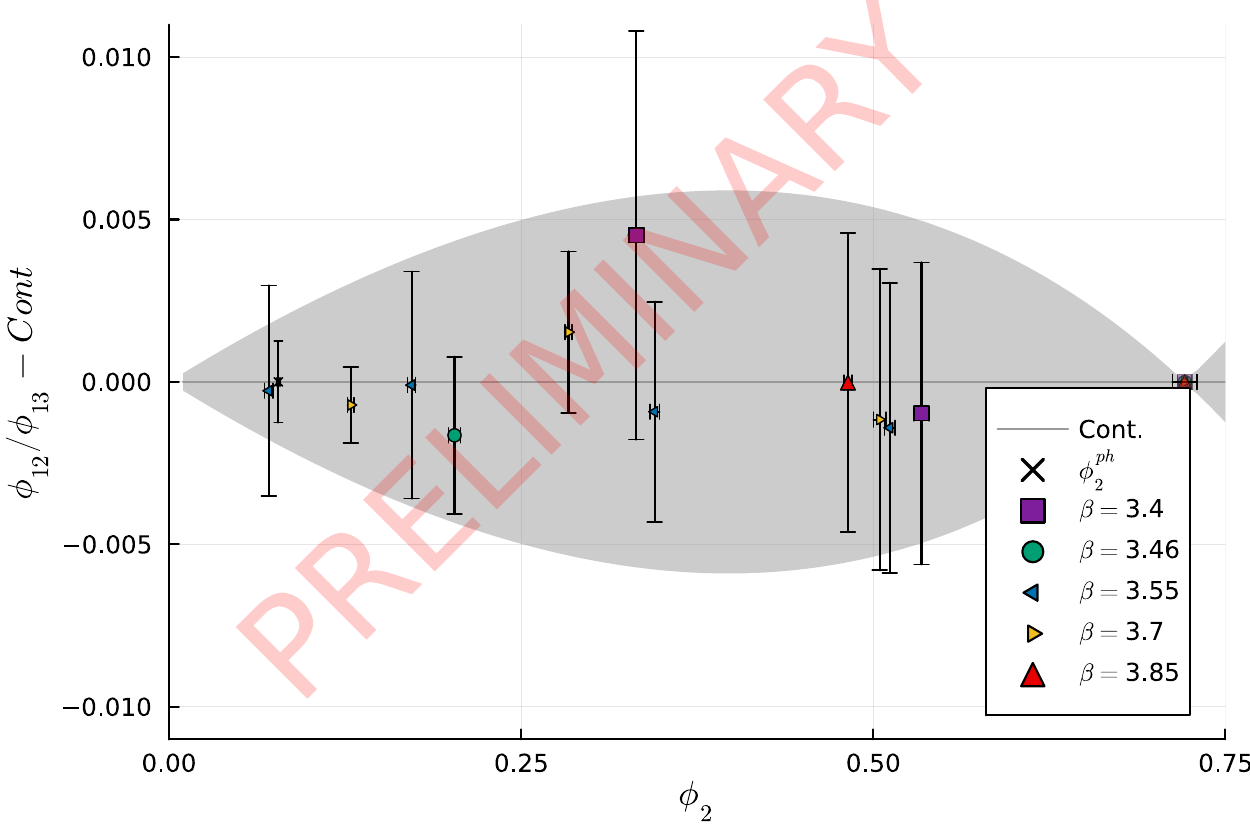}
  \includegraphics[scale=0.35]{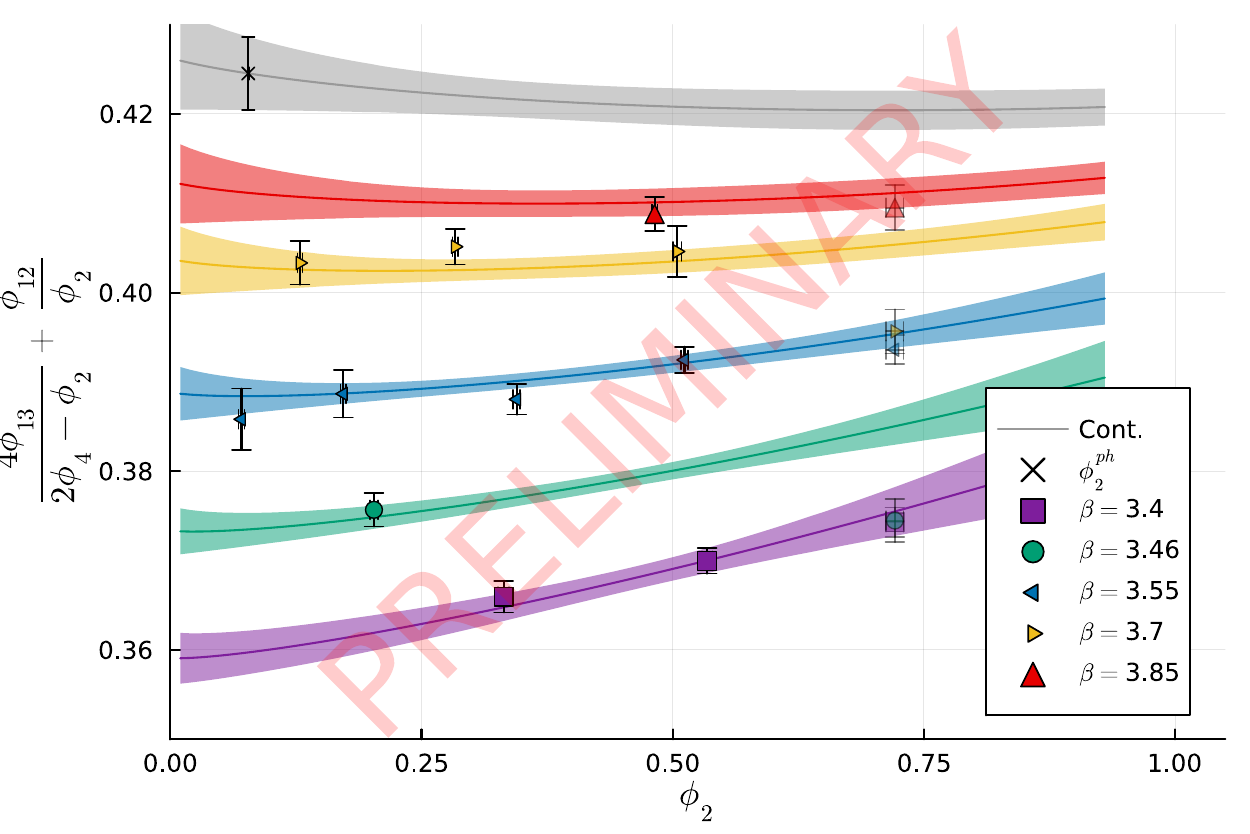}
  \includegraphics[scale=0.35]{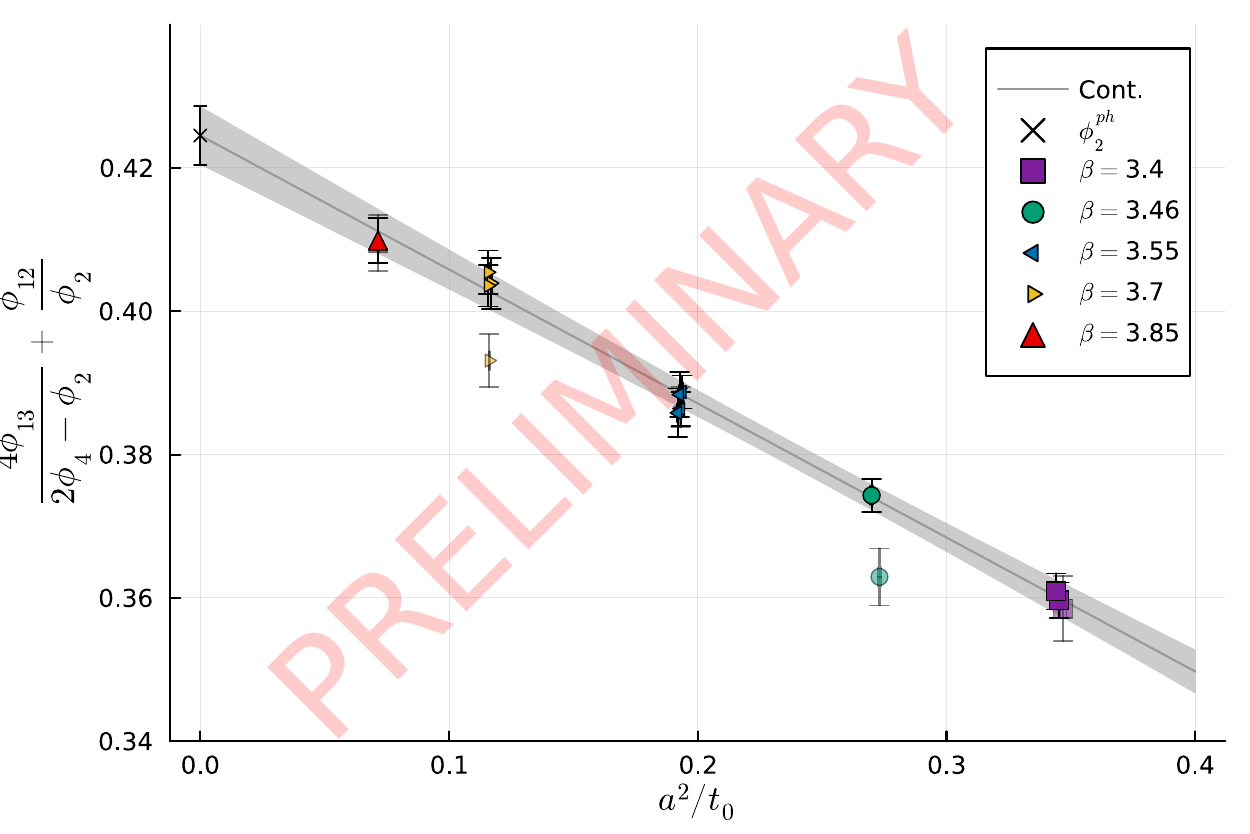}
  \caption{Results for the fit \texttt{[$\chi$ptr][A][$a^{2}\phi$][$m_{\pi} < 420 \textrm{ MeV}$]}, with p-value $\approx$ 0.77. Top row: result for the fit form in eq.(\ref{eq:ratio_1}) plus cutoff effects. The right panel shows the difference with respect to the central value of the continuum curve, without projecting to zero lattice spacing. Bottom row: fit result for eq.(\ref{eq:ratio2}). The left plot show the result for various lattice spacings as a function of the dimensionless pion mass, as well as the continuum extrapolation (gray band). In the right plot we show the continuum-limit scaling, where the data has been projected to physical pion mass using the result of the fit.}
  \label{fig:fits}
\end{figure}

We illustrate the result of these fit functions in Figure \ref{fig:fits}, see section \ref{sec:cutoff} for details on cutoff effects and ansatze notation. %
In the upper plots, we see how the ratio of quark masses is mostly independent of the lattice spacing. %
Notice that, in the upper right plot, the data is not projected to zero lattice spacing, showing that cutoff effects are well below the percent level correction. %
The continuum shape of the fit and the clustering of points at the symmetric point is an effect of the fact that both masses are exactly the same at this point for all lattice spacings.
In the plots on the second row, we see how the chiral effects are suppressed for this quantity, leading to a mostly flat extrapolation to the physical pion mass at zero lattice spacing. %
The points in the last plot are projected to physical pion mass using the result of the fit.

\subsection{Cutoff effects and data cuts}
\label{sec:cutoff}

To parametrize the continuum extrapolation, we model the cutoff effects in the following way:
\begin{itemize}
  \item \texttt{$[a^2]$} : Cutoff effects proportional to $a^{2}/t_{0}$.
  \item \texttt{$[a^2\phi]$} : Massive cutoff effects proportional to $\phi_{2} a^{2}/t_{0}$ or to $\phi_{K} a^{2}/t_{0}$.
\end{itemize}

All these cutoff effects are included in such a way that they are identical for both masses at the symmetric point, i.e., when $\phi_{2}=\phi_{K}$.

\newpage

In order to properly assess systematics, we fit different sets of data labeled by the following tags:
\begin{itemize}
  \item \texttt{[A]} : The mass in eq. (\ref{eq:plat_mq}) is extracted in a Euclidean time fit interval, where deviations from the constant behaviour are smaller then the statistical uncertainty.
  \item \texttt{[B]} : The PCAC mass is extracted through a model average procedure \cite{Frison:2023lwb} (see below) considering a constant fit over a variety of Euclidean time intervals.
  \item \texttt{[$m_{\pi} < X \textrm{ MeV}$]} : We keep ensembles with pion masses below a given value.
  \item \texttt{[$\beta > \beta_{0} $]} : We remove the ensembles at coarsest or second-to-coarsest lattice spacing, leaving only those with $a \leq 0.075 \textrm{ fm}$ or $a \leq 0.063 \textrm{ fm}$.
\end{itemize}
\begin{figure}[htbp]
  \centering
  \includegraphics[scale=0.5]{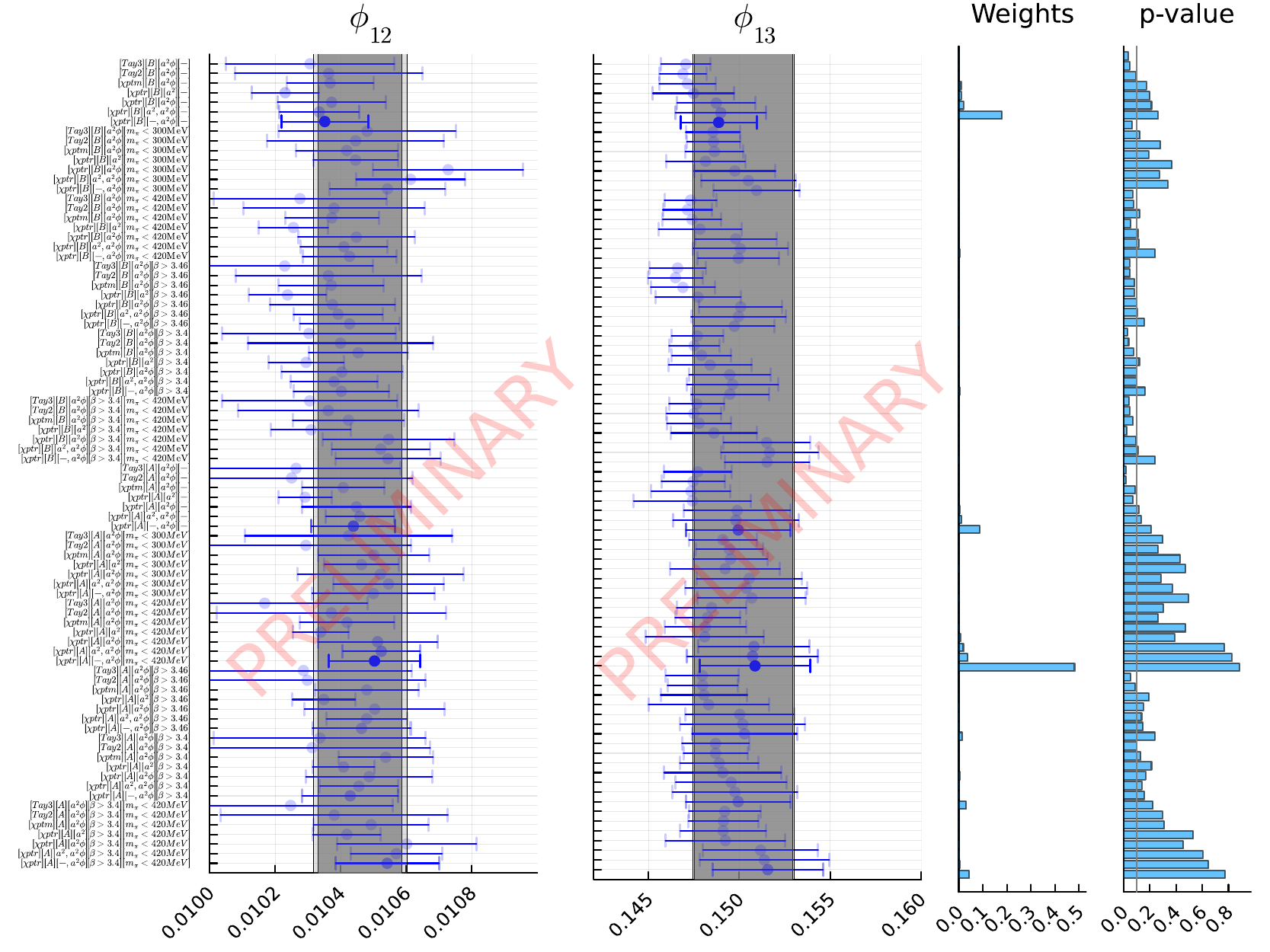}
  \caption{Results for the model average for the considered set of models. The first two plots illustrate the results for the different fits, while the other two plots are the assigned weight and the p-value of the fit (vertical line indicates p-value = 0.1). The opacity of the points is proportional to the weight. The central band is the final result of the model average. The inner darker band corresponds to the statistical error and the lighter outer band to the systematic error from the model averaging procedure.}
  \label{fig:modelaverage}
\end{figure}

We then implement a Model Averaging procedure \cite{Jay:2020jkz,Frison:2023lwb} to obtain a final result and consistent estimation of systematic errors. %
We employ the Takeuchi Information Criterion (TIC) \cite{Takeuchi} where each model gets a weight proportional to

\begin{equation}
  W^{(i)} = \mathcal{N} \exp\left\{ \frac{1}{2}(\chi_{i}^{2} - 2\langle\chi_{i}\rangle^{2})\right\},
\end{equation}
where the normalization factor $\mathcal{N}$ is computed so that the weights add up to one. %
With this, one can assign a value and a systematic error to the model average result:
\begin{equation}
  \langle \mathcal{O} \rangle_{MA} = \sum_{i} W^{(i)}\mathcal{O}^{(i)} ,\qquad \sigma_{syst}^{2} = \langle \mathcal{O}^{2} \rangle_{MA} - \langle \mathcal{O} \rangle_{MA}^{2}.
\end{equation}

After performing the model average and using eq.(\ref{eq:RGI_mass}), we can compute the RG-invariant scheme-independent quark masses. %
While these masses contain all the relevant physical information, it is useful for phenomenological studies to use quark masses in the $\overline{\rm MS}$ scheme at 2 \textrm{GeV} referred to four flavour QCD. %
For this, we can run with four-loop perturbative running to the charm scale with $N_{f} = 3$, and then back to 2 \textrm{GeV} with $N_{f}=4$. %
This process has an effect of around $0.2\%$ with respect to running simply with three flavours.
In order to translate the values of the masses to \textrm{MeV}, we use the scale setting from \cite{Bussone:2025wlf}.

\begin{figure}[h!]
  \centering
  \includegraphics[scale=0.35]{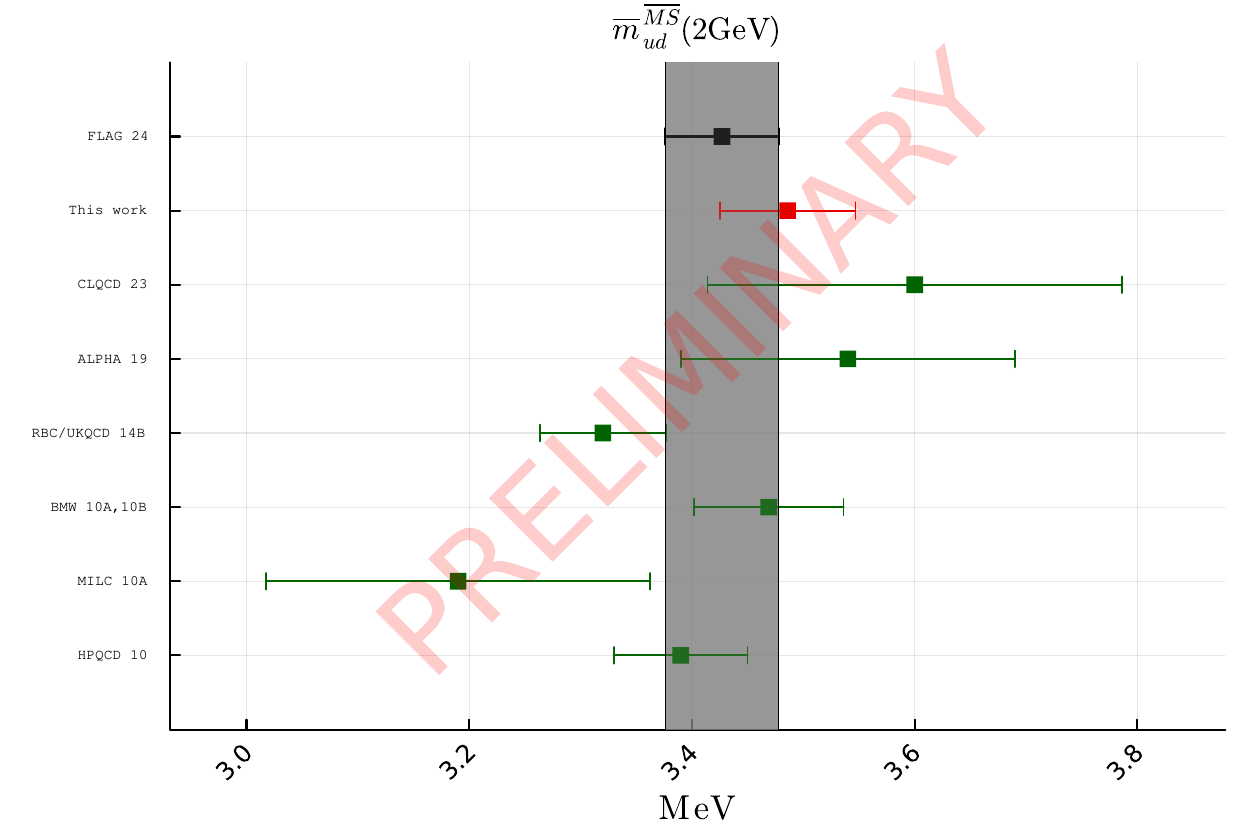}
  \includegraphics[scale=0.35]{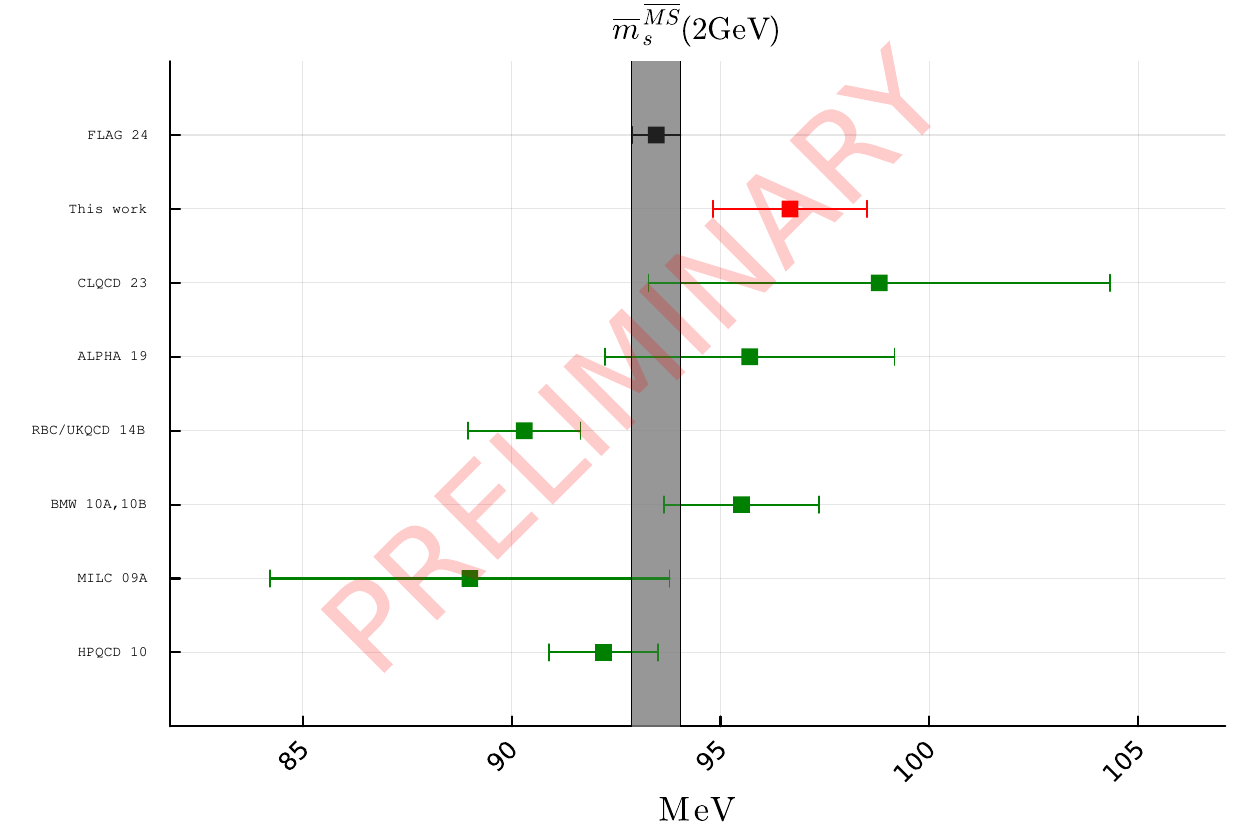}
  \caption{Comparison of light (left) and strange (right) quark masses renormalized in $\overline{MS}$ scheme at 2 \textrm{GeV} in four flavour QCD. The central band corresponds to the FLAG average \cite{FLAG,CLQCD:2023sdb, Bruno:2019vup, RBC:2014ntl, BMW:2010ucx, Bazavov:2010yq, McNeile:2010ji}.}
  \label{fig:flaglike}
\end{figure}

In figure \ref{fig:flaglike} we show a comparison of this work with the FLAG average and the determinations entering that average \cite{CLQCD:2023sdb, Bruno:2019vup, RBC:2014ntl, BMW:2010ucx, Bazavov:2010yq, McNeile:2010ji}.

\begin{table}
  \centering
  \begin{tabular}{c|cc}
    & $\overline{m}_{ud}^{\overline{MS}}(2\textrm{GeV})$ &  $\overline{m}_{s}^{\overline{MS}}(2\textrm{GeV})$\\  \hline
    Relative precision                  & 1.8\% & 1.9 \% \\  \hline
    Model average systematics           & 14\% & 10\%\\
    Statistical + extrapolation         & 33\% & 45\%\\
    RGI running (eq. \ref{eq:RGI_mass}) & 30\% & 25\% \\
    $\overline{MS}$ running             & 23\% & 20\% \\

  \end{tabular}
  \caption{Relative precision and relative error contribution to the quark masses coming from different sources.}
\end{table}

\section{Conclusions}

In this work, we have performed a determination of the light and strange quark masses using $N_{f} = 2+1$ CLS ensembles and the non-perturbative renormalization developed by the ALPHA collaboration.
The systematic effects of the chiral and continuum extrapolation are explored using a model averaging procedure. %

After computing the RGI scheme-independent quark masses, we use 4-loop perturbation theory to obtain the quark masses in the more commonly used $\overline{MS}$-scheme at 2\textrm{GeV}. %
This allows for a comparison of this fundamental parameter with the determination by other collaborations, showing good agreement with the FLAG Report.
With this preliminary results, we find a significant reduction of the error, of around 50-60\%, with respect to \cite{Bruno:2019vup}.

In the future, we plan to extend this work with further exploration of systematics effects and a mixed action setup similar to the one used in ref. %
\cite{Bussone:2025wlf}. %
While in this case the matching strategy of the mixed action setup implies that cutoff effects will be similar between both regularizations. This combined strategy will help control the extrapolation to the physical point and the systematic effects of the determination.

\acknowledgments

We are grateful to our colleagues in the Coordinated Lattice Simulations (CLS) initiative for the generation of the gauge field configuration ensembles employed in this study.
We thank N. Husung, A. Conigli and S. Kuberski for very useful discussions.
We acknowledge PRACE for awarding us access to MareNostrum at Barcelona Supercomputing Center (BSC), Spain and to HAWK at GCS@HLRS, Germany. %
The authors thankfully acknowledge the computer resources at MareNostrum and the technical support provided by Barcelona Supercomputing Center (FI-2020-3-0026). %
We thank CESGA for granting access to the FinisTerrae systems. %
This work is partially supported by grants PID2021-127526NB-I00, and PID2024-160152NB-I00, funded by MCIN/AEI/10.13039/501100011033, by “ERDF A way of making Europe”, and by FEDER, UE, as well as by the Spanish Research Agency (Agencia Estatal de Investigación) through grants IFT Centro de Excelencia Severo Ochoa SEV-2016-0597 and No. CEX2020-001007-S, funded by MCIN/AEI/10.13039/501100011033 and national project CNS2022-136005, AEI/MCIU through grant PID2023-148162NB-C21 and ASFAE/2022/020. %
We also acknowledge support from the project H2020-MSCAITN2018-813942 (EuroPLEx), under grant agreement No. 813942, and the EU Horizon 2020 research and innovation programme, STRONG-2020 project, under grant agreement No. 824093. %
This work is partially funded by the European Commission - NextGenerationEU, through Momentum CSIC Programme: Develop Your Digital Talent. %
We acknowledge HPC support by Emilio Ambite, staff hired under the Generation D initiative, promoted by Red.es, an organisation attached to the Spanish Ministry for Digital Transformation and the Civil Service, for the attraction and retention of talent through grants and training contracts, financed by the Recovery, Transformation and Resilience Plan through the European Union’s Next Generation funds.

\bibliographystyle{JHEP}
\bibliography{biblio}

\end{document}